# Patient-centered appraisal of race-free clinical risk assessment

Charles F. Manski
Department of Economics and Institute for Policy Research
Northwestern University

Revised: February 24, 2022

Until recently, there has been a consensus that clinicians seeking to assess patient risks of illness as a prelude to treatment should condition risk assessments on all observed patient covariates with predictive power. The broad idea is that knowing more about patients enables more accurate predictions of their health risks and, hence, better clinical decisions. This consensus has recently unraveled with respect to a specific covariate, namely race. There have been increasing calls for race-free risk assessment, arguing that using race to predict health risks contributes to racial disparities and inequities in health care. In some medical fields, leading institutions have recommended race-free risk assessment.

An important open question is how race-free risk assessment would affect the quality of clinical decisions. Considering the matter from the patient-centered perspective of medical economics yields a disturbing conclusion: Race-free risk assessment would harm patients of all races.

**Building Tension**

The term *personalized medicine* ideally means health care specific to an individual. However, evidence to support complete personalization is rarely available. Hence, the term commonly means care that varies with observed patient characteristics relevant to clinical decisions. President's Council of Advisors on Science and Technology (1) states (p. 7):

> "personalized medicine does not literally mean the creation of drugs or medical devices that are unique to a patient. Rather, it involves the ability to classify individuals into subpopulations that are uniquely or disproportionately susceptible to a particular disease or responsive to a specific treatment."

In practice, clinicians classify patients into groups based on observed health history, screening and diagnostic tests, and demographic attributes. They predict risks of disease development conditional on these patient covariates and use the findings to inform treatment decisions.

A prominent example is the Breast Cancer Risk Assessment (BCRA) Tool of the National Cancer Institute (https://bcrisktool.cancer.gov/). To use the Tool, one inputs information on a woman's reported medical and reproductive history, history of breast cancer among first-degree relatives (mother, sisters, daughters), age, and race. An algorithm yields estimates of the risk of developing breast cancer over the next 5 years and up to age 90.

Another prominent example is the Arteriosclerosis and Cardiovascular Disease (ASCVD) Risk Estimator of the American College of Cardiology (https://tools.acc.org/ascvd-risk-estimator-plus/#!/calculate/estimate/). Here one inputs information on blood pressure, cholesterol levels, medical history, age, sex, and race. An algorithm uses these covariates to predicts the probability that a person will develop ASCVD in the next ten years.

There has long been a consensus that clinicians should condition risk assessments on all easily observed patient covariates with predictive power. Knowing more about patients should enable more accurate predictions of their health risks and, hence, better clinical decisions. Medical economics has formalized this idea by studying clinicians as planners, who seek to make decisions that maximize patient well-being.

The consensus to base clinical risk assessments on all observed patient covariates with predictive power has recently unraveled with respect to a specific covariate, namely race. Writing in the *New England Journal of Medicine*, Vyas, Eisenstein, and Jones (2) call for a general reconsideration of the use of race in risk assessments. These authors document the current use of race as a covariate in algorithms predicting patient outcomes in cardiology, nephrology, obstetrics, urology, oncology, and other fields of medicine. They assert (p. 874):

> "Many of these race-adjusted algorithms guide decisions in ways that may direct more attention or resources to white patients than to members of racial and ethnic minorities. . . . . Given their potential to perpetuate or even amplify race-based health inequities, they merit thorough scrutiny."

Others argue directly for race-free clinical risk assessments, ones that do not use race as predictors of patient outcomes. Cerdeña, Plaisime, and Tsai (3) write in an Editorial in *The Lancet* (p. 1126), "Race should not be used to make inferences about physiological function in clinical practice." Considering the ASCVD Risk Estimator, a Comment in T*he Lancet Digital Health* (4) advocates "Revising the atherosclerotic cardiovascular disease calculator without race."

In some fields, leading institutions have formally recommended race-free risk assessment. A notable case is Delgado *et al*. (5), which presents the recommendations of the National Kidney Foundation-American Society of Nephrology (NKF-ASN) Task Force on Reassessing the Inclusion of Race in Diagnosing Kidney Disease. The Task Force considered the prevailing use of race in computation of estimated glomerular filtration rate(eGFR), a measure of kidney function. It recommended removal of race as a determinant of eGFR, writing (pp.5-6):

> "For U.S. adults (>85% of whom have normal kidney function), we recommend immediate implementation of the CKD-EPI creatinine equation refit without the race variable in all laboratories in the U.S. because it does not include race in the calculation and reporting, includes diversity in its development, is immediately available to all labs in the U.S., and has acceptable performance characteristics and potential consequences that do not disproportionately affect any one group of individuals."

Research related to the Task Force recommendation is documented by Williams, Hogan, and Ingelfinger (6).

Writers recommending race-free risk assessment have argued that using race to predict patient outcomes contributes to racial disparities and inequities in health care. The assertion by Vyas *et* *al.* (2), quoted above, is typical. Cerdeña *et al.* (3) states (p. 1125) that "race-based medicine . . . ., perpetuates health-care disparities." Based on these and similar statements, it has become common to propose race-free risk assessment. A news article in *Science* (7) provides a general discussion of this movement, its alarming title declaring "A Troubled Calculus: Researchers use race to build disease risk assessment tools. Can removing it help resolve medicine's race crisis?"

Writers calling for race-free risk assessment have not studied how it would affect the quality of clinical decisions. Considering the matter from the patient-centered perspective of medical economics yields a disturbing conclusion: Race-free risk assessment would harm patients of all races. I explain here.

**Optimal Clinical Decisions**

Medical economics considers clinicians as planners who choose how to treat the patients under their care. As described in my book *Patient Care under Uncertainty* (8), the presumed objective is to maximize a well-being (aka welfare) function that sums up the benefits and harms of decisions across the relevant patients. When treatment response is individualistic, a clinician views benefits and harms from the perspective of each patient separately. Pauly (9) refers to such a clinician as an "agent" for the patient. Individualistic treatment response means that the care received by one patient does not affect other patients. This assumption is often realistic when considering non-infectious diseases.

Clinicians commonly observe certain covariates for their patients. Studies in medical economics usually assume that a clinician can make accurate probabilistic risk assessments conditional on these covariates. This assumption does not assert that clinicians can predict patient outcomes with certainty. It only calls for accurate probabilistic predictions. A huge body of medical research aims to provide an evidence-based foundation for making such probabilistic predictions.

In this setting, it can be proved mathematically that optimization of patient care has a simple solution. Patients should be divided into groups having the same observed covariates. All patients in such a group should be given the care yielding the highest within-group mean well-being. Thus, it is optimal to differentially treat patients with different observed covariates if different clinical decisions maximize within-group mean well-being. Patients with the same observed covariates should be treated uniformly.

An important associated finding is that information characterizing patients has value. The magnitude of maximum well-being increases as more patient covariates are observed, provided only that optimal decisions for some patients change when group classifications are refined. Thus, observing more patient covariates is useful if the additional information has predictive power in risk assessment that alters optimal care for some patients. Conversely, shrinking the set of covariates used to predict patient outcomes reduces the quality of patient care.

Formally, let (x, z) be patient covariates observed by the clinician, x being covariates routinely used in risk assessment and z being an additional covariate that may or may not be used. Let $P_{xz}$ be the probability that patients with covariates (x, z) will develop a disease of concern and let $P_x$ be the analogous probability for patients with covariates x. Maximum achievable well-being is larger if $P_{xz}$ is used to assess risk than if $P_x$ is used, provided that $P_{xz}$ varies with z sufficiently to make optimal care vary with z.

These findings are used throughout research on medical economics, for example by Phelps and Mushlin (10), Claxton (11), Meltzer (12), Basu and Meltzer (13), and Manski (8, 14).

**Race in Risk Assessment**

From the perspective of patient-centered care, an observable race classification is simply a covariate that may be used to predict disease in risk assessments, in conjunction with other covariates. The ASCVD Risk Estimator exemplifies the practice. Race has been included in this risk assessment method because it has been found to add predictive power. Karmali et al. (15) analyzes how predictions made vary with the multiple covariates used, race among them.

In the notation introduced above, $P_{xz}$ expresses risk assessment using race and $P_x$ expresses race-free risk assessment. Using $P_x$ rather than $P_{xz}$ to predict illness cannot increase patient well-being. To the contrary, it lowers well-being when optimal clinical decisions made using $P_x$ differ from those made using $P_{xz}$.

Recent calls for race-free assessment make assertions that are strikingly at odds with this perspective. One assertion has been that the legitimacy of using race to assess health risks is related to the controversial question of the extent to which race is a social or biological concept. The NKF-ASN Task Force (5) stated (p. 7): "The rationale for the Task Force includes: race is a social and not a biological construct." See also Cerdeña *et al.* (3).

In practice, race may be classified in various ways—by patient selection among classifications, by patient responses to questions about ancestry, or by researcher interpretation of patient characteristics. However race is classified, its inclusion among the covariates used to predict illness does not require that one take any position on the extent to which the classification expresses biological or social characteristics of patients. From the patient-centered perspective, inclusion of race is well-motivated whenever it has predictive power. Predictive power and data availability have long been the criteria used to choose covariates in medical risk assessment.

Consider prediction of breast cancer in the BCRA Tool, using history of breast cancer among first-degree relatives. This covariate may have predictive power in part because relatives may have genetic similarities and in part because relatives may live in similar environments. Justifying use of the medical history of relatives to predict breast cancer does not require one to know how similarities of biology and environment combine to yield predictive power. The same reasoning applies to race.

A second assertion has been that justification for using race as a covariate in risk assessment requires establishment of a causal link between race and illness. Vyas *et al.* (2) wrote (p. 880): "When developing or applying clinical algorithms, physicians should ask . . . Is there a plausible causal mechanism for the racial difference that justifies the race correction?" In fact, causal analysis is unnecessary in patient-centered clinical risk assessment. The concern of risk assessment is statistical association rather than causation.

A third assertion is that using race as a covariate has, in the language of Vyas *et al.* (2) quoted earlier: "potential to perpetuate or even amplify race-based health inequities." Their assertion is unfounded if clinicians behave as patient-centered planners. If so, clinical decision making should not yield inequities in the sense of lack of fairness or justice. The patient-centered objective is to maximize patient well-being, optimizing care within groups of patient who share common observed covariates. Thus, it embeds a clear idea that decision making should be fair and just.

Of course, even fair and just decisions may be imperfect. Risk assessments aim to predict health risk as well as possible using readily observed patient covariates. Prediction accuracy may be improved if it is feasible to observe further covariates that add predictive power. Covariates that are currently used in risk assessment may become superfluous if informative new covariates become widely observable and embedded in assessment algorithms. For example, prediction of ASCVD with findings from CT scans and electrocardiograms might supplant use of blood pressure and cholesterol levels. Prediction of breast cancer with knowledge of patients' DNA and observation of their environments might make it irrelevant to use self-reports on the histories of breast cancer among first-degree relatives. Similarly, knowledge of DNA and observation of patient environments might make it irrelevant to use race as a covariate in risk assessments.

A fourth assertion is that many patients take offense to race being used in clinical risk assessment. I am not aware of research that would enable one to assess what fraction of the American population hold this normative position. Among those who hold the position, I do not know how often persons view it as a moral imperative to refrain from using race in risk assessment, regardless of its predictive usefulness. Some persons may hold the position in a less absolutist way. They may prefer not to use race in risk assessment, all else equal, yet be willing to do so if the health benefits are sufficiently large.

Why might some persons take offense to race being used in risk assessment, even when its use helps to predict patient health? Anecdotal evidence suggests a visceral fear, stemming from the tortured racial history of our society and concerns about medical practices associated with race, that data on race will be misused rather than employed to benefit patients. This fear is understandable. However, I think it would be a serious mistake to permit it to shut down the well-motivated efforts of biostatisticians and epidemiologists to improve clinical risk assessment and the similarly well-motivated desire of clinicians and patients to make risk assessment as accurate as possible.

**Finding a Way Forward**

Recommendations for race-free clinical risk assessment may have laudable intentions. However, considering the matter from the perspective of patient-centered decision making yields the conclusion that implementing these recommendations would yield negative health consequences for patients of all races.

This conclusion does not imply that all is well in the manner that race is currently used in health assessments. In their A*lgorithmic Bias Playbook*, Obermeyer et al. (16) call attention to two potential categories of what they call *algorithmic bias*. (Here bias refers to statistical bias in estimation, not psychological bias in clinical decisions). They write (p. 2):

> "The first is when algorithms are aimed at the right target, but fail to hit it for underserved groups. This is often because they were trained or evaluated in non-diverse populations. . . . . Throughout our work on algorithmic bias, though, we've found that a second category is far more common: algorithms are *aimed at the wrong target* to begin with. The result is an insidious 'label choice bias,' arising from a mismatch between the ideal target the algorithm *should be predicting*, and a biased proxy variable the algorithm *is actually predicting*."

Yet the authors' insightful discussion of these potential biases does not lead them to recommend removal of race as a covariate in clinical risk assessment. They write (p. 11):

> "If the algorithm is predicting its ideal target, we may want the algorithm to use race and zip code variables: they can help it predict the ideal target better. On the other hand, if the algorithm's actual target is a biased proxy, it will be biased *regardless* of the input variables you include. . . . . The key take away: focus on the target variable the algorithm is predicting — not the variables the algorithm uses to predict it."

How then might one justify race-free risk assessment? One possibility would be to perform empirical analysis that seeks to learn whether clinical decision making deviates from patient-centered principles in ways that perpetuate or amplify racial inequities. Such research may be useful, but I expect that it will be difficult to reach clear conclusions. Although (2) use the term "race based" when describing clinical risk assessments that use race as a covariate, these assessments are not based on race alone. Evidence-based assessment tools provide clinicians with estimates of illness probabilities of the form $P_{xz}$, which condition prediction on all the covariates $(x, z)$, not just z. Hence, the predictions that clinicians use to make decisions are based on many patient characteristics, not only race. A further complication for empirical analysis is that the non-racial covariates used in risk assessment are often statistically associated with racial classification.

Another possibility is to argue that the patient-centered perspective misses fundamental societal concerns when it views race as simply one of many patient covariates used in risk assessment. If an alternative perspective is to have a compelling foundation, it should explain why society should find it acceptable to make risk assessments using other patient characteristics that clinicians observe, but not race. It should explain why the social benefit of omitting race from risk assessment is sufficiently large that it exceeds the harm to the quality of patient care.